\documentclass[fleqn]{article}
\usepackage{favorite_preamble}

\title{
	\LARGE \textbf{
		Gravitational formfactors of the $ \rho $ meson in QCD sum rules
	}
}
\author{
	T. M. Aliev\thanks{email: \href{mailto:taliev@metu.edu.tr}{taliev@metu.edu.tr}}\\ \textit{\normalsize Physics Department, Middle East Technical University, Ankara 06800, Turkey} \\$  $\\
	T. Barakat\thanks{email: \href{mailto:tbarakat@ksu.edu.sa}{tbarakat@ksu.edu.sa}}\\ \textit{\normalsize Physics Department, King Saud University, Riyadh 11451, Saudi Arabia} \\$  $\\
	K. \c Sim\c sek\thanks{email: \href{mailto:ksimsek@ur.rochester.edu}{ksimsek@ur.rochester.edu}}\\ \textit{\normalsize Department of Physics \& Astronomy, University of Rochester, Rochester, NY 14627, USA}
}
\date{\today}

\newcommand{\frho}{f _\rho}
\newcommand{\mrho}{m _\rho}
\newcommand{\gpera}{g _\perp ^a (u)}
\newcommand{\gperv}{g _\perp ^v (u)}
\newcommand{\phipar}{\phi _\parallel (u)}
\newcommand{\Apar}{A _\parallel (u)}
\newcommand{\guc}{g _3 (u)}
\newcommand{\gperap}{{g _\perp ^a}' (u)}

\newcommand{\hgperv}{\hat g _\perp ^v (u)}
\newcommand{\hphipar}{\hat \phi _\parallel (u)}
\newcommand{\hApar}{\hat A _\parallel (u)}
\newcommand{\hguc}{\hat g _3 (u)}

\newcommand{\hhgperv}{\hat{\hat g} _\perp ^v (u)}
\newcommand{\hhphipar}{\hat{\hat\phi} _\parallel (u)}

\newcommand{\hhguc}{g _3 (u)}
\newcommand{\I}{\mathcal I}
\newcommand{\V}{\mathcal V(\alpha _i)}
\newcommand{\A}{\mathcal A(\alpha _i)}

\begin{document}
\maketitle
\begin{abstract}
	By using the quark part of the energy-momentum tensor current, the gravitational formfactors of the $ \rho $ meson are calculated within the light-cone sum rules method. In the considered version, the energy-momentum tensor current is not conserved and as a result, there appear nine formfactors, six (three) of which correspond to the conservation (nonconservation) of the energy-momentum tensor current. We also compare our results with the one existing in the literature.
\end{abstract}
\section{Introduction}
During the last years, the energy-momentum tensor (EMT) has become one of the popular research objects for better understanding the structure of hadrons \cite{ref:1,ref:2}. The gravitational formfactors (GFFs) are defined with the help of the matrix element of the symmetric EMT (see \cite{ref:3, ref:4}). The GFF relates the mass, spin, total angular momentum, etc. Understanding the EMT can help answer questions as to the origin of the nucleon mass, spin carried by quarks and gluons, and how the strong force is distributed inside hadrons. The total GFFs were introduced a long time ago for both spin-0 and spin-1/2 hadrons \cite{ref:5}. The GFFs for spin-1 particles have been discussed in the literature \cite{ref:6,ref:7,ref:8,ref:9,ref:10}. 
\prg 
The GFFs of the $ \rho $ meson was studied within the light-cone constituent quark model in \cite{ref:11}. It should be emphasized that the study of GFFs for hadrons started with \cite{ref:Ji.1} and experimental measurements through deeply virtual Compton scattering \cite{ref:Ji.2}. The GFFs were also calculated in chiral perturbation in \cite{ref:Ji.3}.
\prg
In the present work, we study the GFFs of the $ \rho $ meson within the light-cone sum rules. The paper is organized as follows. In Section 2, we derive the sum rules for the GFFs under study. In Section 3, we perform the numerical analysis for the sum rules obtained in the previous section. Section 4 contains our conclusion.

\section{Sum rules for the $ \rho $ meson GFFs}
For the calculation of the GFFs of the $ \rho $ meson within the light-cone sum rules, we introduce the following correlation function:
\begin{align}
\Pi _{\mu\nu\lambda} &= i \int d^4x \ e^{iqx} \bracket{\rho (p)}{T \{T _{\mu\nu}^q (x) J _\lambda (0)\}}{0}
\end{align}
where $ J _\lambda = \bar u \gamma _\lambda d $ is the interpolating current of the $ \rho $ meson, the $ T _{\mu\nu}^q (x) $ is the EMT including only the contribution of quark fields,
\begin{align}
T _{\mu\nu} ^q &= \frac i4 \Big[
\bar \psi (\doublevec{\mathcal D} _\mu \gamma _\nu + \doublevec{\mathcal D} _\nu \gamma _\mu) \psi - g _{\mu\nu} \frac i2 (\doublevec{\slashed{\mathcal D}} - m _q) \psi 
\Big]
\end{align}
where $ \doublevec{\mathcal D} _\mu = \doublevec{\del} _\mu \pm i g A _\mu ^a \frac{\lambda ^a}{2} $. 
\prg
In \cite{ref:3}, it is obtained that the second term of the EMT can be written as $ g _{\mu\nu} (\doublevec{\slashed{\mathcal D}} - m _q) \psi \approx g _{\mu\nu} ( 1 + \gamma _m ) m _q  q \bar q  $ where $ \gamma _m $ is the anomalous dimension of the mass operator. In the present work, we are working in the chiral limit ($ m _q \to 0$), hence the second term of the EMT can be neglected.
\prg 
We start our consideration by computing the correlation function from the hadronic side. Its representation from the hadronic side is obtained by inserting a complete set of mesons carrying the same quantum numbers as the $ \rho $ meson and, isolating the contribution of the ground state, we get
\begin{align}
\Pi _{\mu\nu\lambda} &= \frac{\bracket{\rho (p)}{T _{\mu\nu} ^q }{\rho (p')}\bracket{\rho (p')}{J _\lambda}{0}}{p'^2 - m _\rho^2} + \mbox{higher states} \label{2}
\end{align}
The matrix elements in Eq. \eqref{2} are defined as
\begin{align}
\bracket{\rho (p')}{J _\lambda}{0 } &= f _\rho m _\rho \epsilon _\lambda ^* 
\end{align}
where $ f _\rho $ is the $ \rho $ meson decay constant, $ m _\rho $ is its mass, and $ \epsilon _\lambda $ is its polarization vector. The EMT of a spin-1 particle in QCD is defined as (see, for example, \cite{ref:10, ref:10.1})
	\begin{align}
	\bracket{\rho (p)}{T _{\mu\nu}^q}{\rho (p')} &= 2 P _\mu P _\nu \Big(
	- \epsilon '^* \cdot \epsilon A _0 (q^2) + \frac{\epsilon'^* \cdot P \epsilon \cdot P}{m _\rho ^2} A _1 (q^2)
	\Big) +\nn 
	+ 2 [P _\mu (\epsilon'^* _\nu \epsilon \cdot P + \epsilon _\nu \epsilon'^* \cdot P) + P _\nu (\epsilon'^* _\mu \epsilon \cdot P + \epsilon _\mu \epsilon'^* \cdot P)] J (q^2) \nn 
	+ \frac 12 (q _\mu q _\nu - g _{\mu\nu} q^2) \Big(
	\epsilon'^* \cdot \epsilon D _0 (q^2) + \frac{\epsilon'^* \cdot P \epsilon \cdot P}{m _\rho ^2} D _1(q^2) 
	\Big) \nn 
	+ \Big[
	\frac 12 (\epsilon _\mu \epsilon'^* _\nu + \epsilon'^* _\mu \epsilon _\nu) q^2 + (\epsilon '^* _\mu q _\nu + \epsilon '^* _\nu q _\mu) \epsilon \cdot P - (\epsilon _\mu q _\nu + \epsilon _\nu q _\mu) \epsilon '^* \cdot P 
	- 4 g _{\mu\nu} \epsilon'^* \cdot P \epsilon \cdot P
	\Big] E (q^2) \nn 
	+ \Big(
	\epsilon _\mu \epsilon '^* _\nu + \epsilon '^* _\mu \epsilon _\nu - \frac 12 \epsilon '^* \cdot \epsilon g _{\mu\nu}
	\Big) m _\rho^2 F(q^2)  
	+ g _{\mu\nu} (\epsilon '^* \cdot \epsilon m _\rho ^2 C _0 (q^2) + \epsilon '^* \cdot P \epsilon \cdot P C _1(q^2)) \label{4}
	\end{align}
where $ P = \frac 12 (p+p') $ and $ q = p'-p $. Here, the superscript $ q $ indicates that the considered GFF contains contributions only from the quark part of the EMT. 
\prg
In Eq. (5), the first six formfactors are individually (separately for quark and gluon fields) energy-momentum conserving and the remaining three are not. Due to the sum of quark and gluon parts, the EMT conservation leads to the constraints $ \sum _{a = q,g} F ^a(Q^2) = \sum _{a=q,g} C _0 ^a = \sum _{a=q,g} C _1 ^a = 0  $. In addition to these constraints, there are the normalization conditions $ \sum _{a = q,g} A _0^a (Q^2) = 1 $ and $ \sum _{a = q,g} J ^a(Q^2) = 1 $.
\prg  
Since in Eq. \eqref{4} we have nine formfactors, we need nine independent Lorentz structures for the determination of these formfactors. We choose the following structures:
\begin{align}
\Pi _{\mu\nu\lambda} &= 
\epsilon '^* _\lambda p _\mu q _\nu \Pi _1
+ p _mu g _{\mu\nu} \epsilon'^* \cdot q \Pi _2
+ \epsilon '^* _\mu p _\lambda q _\nu \Pi _3
 + \epsilon '^* _\lambda q _\mu q _\nu \Pi _4 
+ p _\nu p _\lambda q _\mu \epsilon'^* \cdot q \Pi _5
+ q _\mu q _\nu p _\lambda \epsilon'^* \cdot q \Pi _6
\nn + \epsilon '^* g _{\mu\nu} \Pi _7
+ p _\lambda g _{\mu\nu} \epsilon'^* \cdot q \Pi _8
+ q _\mu g _{\nu\lambda} \epsilon'^* \cdot q \Pi _9
 + \cdots 
\end{align}
Now, we calculate the correlation function from the QCD side. Using the explicit forms of the interpolating currents $ J _\lambda $ and $ T _{\mu\nu}^q $ and by using the Wick theorem for the theoretical part of the correlation function, we get
\begin{align}
\Pi _{\mu\nu\lambda} &= -\frac 14 \int d^4 x\ e^{iqx} \langle \rho (p') \vert [
\bar u (x) \gamma _\mu \doublevec{\mathcal D} _\nu S (x) \gamma _\lambda d (0)
 + \bar u(0) \gamma _\lambda \doublevec{\mathcal D} _\mu S (x) \gamma _\nu d (x)
+ \bar u (x) \gamma _\nu \doublevec{\mathcal D} _\mu S (x) \gamma _\lambda d (0)
\nn + \bar u(0) \gamma _\lambda \doublevec{\mathcal D} _\nu S (x) \gamma _\mu d (x) ]
\vert 0\rangle 
\end{align}
From this formula, it follows that for the calculation of the correlation function from the QCD side, the expression of the quark propagator in the presence of a background field is needed. This expression was obtained in \cite{ref:13} as
\begin{align}
S _q (x) &= \frac{i\slashed x}{2\pi^2 x^4} - \frac{ig _s}{16\pi^2 x^2} \int _0^1 du\ [\bar u \sigma _{\alpha\beta} \slashed x + \slashed x \sigma _{\alpha\beta} u ] G ^{\alpha\beta} 
 -  \frac{i e _q}{16\pi^2 x^2} \int _0 ^1 du \ [\bar u \sigma _{\alpha\beta} \slashed x + \slashed x \sigma _{\alpha\beta}] F ^{\alpha\beta} 
\end{align}
where $ G ^{\alpha\beta} $ and $ F ^{\alpha\beta} $ are the gluon and photon field strength tensors, respectively. Performing relevant calculations for the correlation function from QCD, we get
\begin{align}
\Pi _{\mu\nu\lambda} &= - \frac 14 \Big\{
\frac{i}{2\pi^2} \sum _i \int du \Big[
\frac 14 \bracket{\rho}{\bar u \Gamma _i d}{0}   \tr \Big\{ \Gamma _i \gamma _\nu \Big(
\frac{\gamma _\mu}{x^4} - \frac{4x _\mu \slashed x}{x^6}
\Big) \gamma _\lambda \Big\} + \frac 14 \bracket{\rho}{\bar u \Gamma _i d}{0} \nn \times \tr \Big\{ \Gamma _i \gamma _\lambda \Big(
\frac{\gamma _\mu}{x^4} - \frac{4x _\mu \slashed x}{x^6} 
\Big) \gamma _\nu \Big\} + (\mu \leftrightarrow \nu)
\Big]  - \frac{i g _s}{16\pi ^2} \sum _i \int du\ \Big[
\frac14 \bracket{\rho}{\bar u \Gamma _i G ^{\alpha\beta} q}{0} \nn \times \tr \Big\{
\Gamma _i \gamma _\lambda \Big(
\Big(
\frac{\gamma _\mu}{x^2} - \frac{2x _\mu \slashed x}{x^4} 
\Big) \bar u \sigma _{\alpha\beta}  +  u \sigma _{\alpha\beta} \Big(
\frac{\gamma _\mu}{x^2} - \frac{2x _\mu \slashed x}{x^4}
\Big) 
\Big) \gamma _\nu
\Big\} + \frac 14 \bracket{\rho}{\bar u \Gamma _i G ^{\alpha\beta} q}{0} \nn\times \tr \Big\{
\Gamma _i \gamma _\nu \Big(
\bar u \Big(
\frac{\gamma _\mu}{x^2} - \frac{2x _\mu \slashed x}{x^4} 
\Big)\sigma _{\alpha\beta}  + u \sigma _{\alpha\beta} \Big(
\frac{\gamma _\mu}{x^2} - \frac{2x _\mu \slashed x}{x^4}
\Big)
\Big) \gamma _\lambda
\Big\} + (\mu \leftrightarrow \nu)
\Big] \Big\}
\end{align}
In these expressions, $ \Gamma _i = \{1, \gamma _5, \gamma _\mu, i \gamma _\mu \gamma _5, \frac{1}{\sqrt 2} \sigma _{\mu\nu}\} $ is the full set of Dirac matrices. The matrix elements $ \bracket{\rho}{\bar u \Gamma _i d}{0} $ and $ \bracket{\rho}{\bar u \Gamma _i G _{\alpha\beta} d}{0} $ are expressed in terms of the $ \rho $ meson distribution amplitudes (DAs) of different twists. These DAs are the main nonperturbative parameters of the light-cone sum rules. Of the aforementioned matrix elements, only the vector and axial components survive after taking the trace, which are defined as (see \cite{ref:14, ref:15, ref:15.1, ref:16})
	\begin{align}
	\bracket{\rho (p)}{\bar u (x) \gamma _\mu d (0)}{0} &= f _\rho m _\rho \Big\{
	\frac{\epsilon '^* \cdot x}{p\cdot x} p _\mu \int _0 ^1 du \ e^{i\bar u px}\Big[
	\phi _\parallel (u) + \frac{m _\rho ^2 x^2}{16} A _\parallel (u)
	\Big] 
	\nn + \Big(
	\epsilon '^* _\mu - p _\mu \frac{\epsilon'^* \cdot x}{p\cdot x} \int _0 ^1 du \ e^{i\bar upx} g _\perp ^v (u)
	\Big)
	\nn - \frac 12 x _\mu \frac{\epsilon'^* \cdot x}{(p\cdot x)^2} m _\rho ^2 \int _0^1 du \ e^{i\bar u px} \Big[
	g _3 (u) + \phi _\parallel (u) - 2 g _\perp ^v (u)
	\Big]
	\Big\}\\
	\bracket{\rho (p)}{\bar u (x) i \gamma _\mu \gamma _5 d (0)}{0} &= - \frac i4 \varepsilon _{\mu\nu\alpha\beta} {\epsilon'^*}^\nu p^\alpha x^\beta f _\rho m _\rho \int _0 ^1 du \ e^{i\bar u px} g _\perp ^a (u)\\
	\bracket{\rho (p)}{\bar u (x) G _{\alpha\beta} \gamma _\mu d (0)}{0} &= \frac{f _\rho m _\rho}{ig _s} p _\mu (\epsilon'^* _\alpha p _\beta - \epsilon'^* _\beta p _\alpha) \int \mathscr D\alpha _i \ e^{i(\alpha _1 + \bar u \alpha _3)px} \mathcal V (\alpha _i)\\
	\bracket{\rho (p)}{\bar u (x) \tilde G _{\alpha\beta} i \gamma _\mu \gamma _5 d(0)}{0} &= \frac{i f _\rho m _\rho}{g _s} p _\mu (\epsilon'^* _\alpha p _\beta - \epsilon'^* _\beta p _\alpha) \int \mathscr D \alpha _i \ e^{i(\alpha _1 + \bar u \alpha _3) px} \mathcal A (\alpha _i)
	\end{align}
where $ \tilde G _{\alpha\beta} = \frac 12 \varepsilon _{\alpha\beta\mu\nu} G ^{\mu\nu} $ is the dual gluon field strength tensor and $ \int \mathscr D \alpha _i = \int d\alpha _1\ d\alpha _2\ d\alpha _3\ \delta (1-\alpha _1 - \alpha _2 - \alpha _3) $. Using the DAs and performing Fourier and Borel transformations on the theoretical part of the correlation function, we get
	\begin{align}
	\Pi _1 &= \frac 18 \frho \mrho (\I_1[\gpera+(\gperap+4 \gperv) (-1+u),1]+\I_3[\gpera+\gperap u-4 \gperv u,1]
	\nn +4 \mrho^2 (2 \I_1[(-2 \hgperv+\hguc+\hphipar) (-1+u),2]-2 \I_3[(-2 \hgperv+\hguc+\hphipar) u,2]
	\nn +\I_5[(\alpha_1+\alpha_3-\alpha_3 u) (-\V+\A (-1+2 u)),2]-\I_6[(\alpha_1+\alpha_3 u) (-\V+\A (-1+2 u)),2]))\\
	\Pi _2 &= \frac 18 \frho \mrho (4 (\I_1[\hgperv-\hphipar+\phipar-\phipar u,1]+\I_3[-\hgperv+\hphipar+\phipar u,1])
	\nn +\mrho^2 (-\I_1[\hApar+2 \Apar (-1+u)+4 (-2 \hgperv+\hguc+\hphipar) (-1+u),2]
	\nn +\I_3[\hApar+2 \Apar u+4 (-2 \hgperv+\hguc+\hphipar) u,2]+4 (\I_5[(\alpha_1+\alpha_3-\alpha_3 u) (\A+\V-2 \V u),2]
	\nn +\I_6[(\alpha_1+\alpha_3 u) (\A+\V-2 \V u),2])))\\
	\Pi _3 &= \frac 14 \frho \mrho (2 \I_1[\hgperv-\hphipar,1]+2 \I_3[-\hgperv+\hphipar,1]
	\nn +\mrho^2 (-\I_1[\hApar-4 (-2 \hgperv+\hguc+\hphipar) (-1+u),2]+\I_3[\hApar-4 (-2 \hgperv+\hguc+\hphipar) u,2]))\\
	\Pi _4 &= \frac 18 \frho \mrho (-\I_1[\gperap+4 \gperv,1]+\I_3[\gperap-4 \gperv,1]+4 \mrho^2 (-2 \I_1[-2 \hgperv+\hguc+\hphipar,2]
	\nn -2 \I_3[-2 \hgperv+\hguc+\hphipar,2]+\I_5[-\V+\A (-1+2 u),2]+\I_6[\A+\V-2 \A u,2]))\\
	\Pi _5 &= \frho \mrho (2 \I_1[(\hgperv-\hphipar) (-1+u),2]+2 \I_3[(\hgperv-\hphipar) u,2]
	\nn +2 \mrho^2 (\I_1[(\hApar-2 (-2 \hgperv+\hguc+\hphipar) (-1+u)) (1-u),3]
	\nn +\I_3[u (-\hApar+2 (-2 \hgperv+\hguc+\hphipar) u),3])-\I_5[\A+\V-2 \V u,2]
	\nn -\I_6[\A+\V-2 \V u,2])\\
	\Pi _6 &= \frac 14 \frho \mrho (-\I_1[\gpera+4 \hgperv-4 \hphipar,2]-\I_3[\gpera-4 \hgperv+4 \hphipar,2]
	\nn +4 \mrho^2 (\I_1[\hApar-4 (-2 \hgperv+\hguc+\hphipar) (-1+u),3]-\I_3[\hApar-4 (-2 \hgperv+\hguc+\hphipar) u,3])) \\
	\Pi _7 &= -\frac 18 \frho \mrho (\mrho^2 (\I_1[\Apar+2 (-2 \hgperv+\hguc+\hphipar),1]+\I_3[\Apar+2 (-2 \hgperv+\hguc+\hphipar),1])
	\nn -2 (\I_2[-\gperv+\phipar]+\I_4[-\gperv+\phipar]))\\
	\Pi _8 &= \frac 18 \frho \mrho (\I_1[\gpera,1]+\I_3[\gpera,1]+\mrho^2 (-\I_1[\hApar-4 (-2 \hgperv+\hguc+\hphipar) (-1+u),2]
	\nn +\I_3[\hApar-4 (-2 \hgperv+\hguc+\hphipar) u,2])) \\
	\Pi _9 &= \frac 14 \frho \mrho (2 (\I_1[\phipar,1]+\I_3[\phipar,1])+\mrho^2 (\I_1[\Apar+2 (-2 \hgperv+\hguc+\hphipar),2]
	\nn +\I_3[\Apar+2 (-2 \hgperv+\hguc+\hphipar),2]+2 (\I_5[\A+\V-2 \V u,2]
	\nn +\I_6[\A+\V-2 \V u,2])))
	\end{align}
where the hat denotes integration, for example, as $ \hat f (u) = \int _0 ^u dv\ f(v) $, and the functions $ \mathcal I _i [f(u),n] $, $ \mathcal I _j [f (u)] $, and $ \mathcal I _k [f(u)\mathcal F (\alpha _i),n ] $ are defined as 
	\begin{align}
	\mathcal I _1 [f(u), n] &= (-1)^n \int _0 ^{u _{10}} du \ \frac{F _{1n}(u)}{(n-1)! (M^2)^{n-1}} e^{-s _1(u)/M^2} 
	\nn - \Big[
	\frac{(-1)^{n-1}}{(n-1)!} e^{-s _1(u)/M^2} \sum _{\ell = 1}^{n-1} \frac{1}{(M^2)^{n-\ell - 1}} \frac{1}{s _1' (u)} \Big(\frac{d}{du} \frac{1}{s _1'(u)}\Big) ^{\ell -1} F _{1n} (u) 
	\Big] _{u=u _{10}} \label{I1}\\
	\mathcal I _2 [f(u)] &= - \int _0 ^{u _{20}} du \ f(u) M^2 e^{-s _2(u)/M^2}\\
	\mathcal I _3 [f(u),n] &= (-1)^n \int _0 ^{u _{30}} du \ \frac{F _{3n}(u)}{(n-1)! (M^2)^{n-1}} e^{-s _3(u)/M^2} 
	\nn- \Big[
	\frac{(-1)^{n-1}}{(n-1)!} e^{-s _3(u)/M^2} \sum _{\ell = 1}^{n-1} \frac{1}{(M^2)^{n-\ell - 1}} \frac{1}{s _3' (u)} \Big(\frac{d}{du} \frac{1}{s _3'(u)}\Big) ^{\ell -1} F _{3n} (u) 
	\Big] _{u=u _{30}} \label{I3} \\
	\mathcal I _4 [f(u)] &= - \int _0 ^{u _{40}} du \ f(u) M^2 e^{-s _4(u)/M^2}\\
	\mathcal I _5 [f(u)\mathcal F(\alpha _i), n] &= (-1)^n \int _0 ^1 du \int _0^{y _{0}} d\alpha _1 \int _{(y _{0}-\alpha _1)/\bar u} ^{1-\alpha _1} d\alpha _3 \ \frac{f(u) \tilde{\mathcal F} _5 (\alpha _i)}{(n-1)! (M^2)^{n-1}} e^{-s _5(u)/M^2}\\
	\mathcal I _6 [f(u)\mathcal F(\alpha _i), n] &= (-1)^n \int _0 ^1 du \int _0^{y _{0}} d\alpha _1 \int _{(y _{0}-\alpha _1)/u} ^{1-\alpha _1} d\alpha _3 \ \frac{f(u) \tilde{\mathcal F} _6(\alpha _i)}{(n-1)! (M^2)^{n-1}} e^{-s _6(u)/M^2}
	\end{align}
with 
\begin{align}
s _1 (u) &= s _2 (u) = m _\rho ^2 u - \frac{u}{\bar u }q^2\\
s _3 (u) &= s _4 (u) = m _\rho ^2 \bar u - \frac{\bar u}{u }q^2\\
s _5 (u) &= s _6 (u) = m _\rho ^2 \bar y - \frac{\bar y}{y} q^2
\end{align}
and
\begin{align}
F _{1n} (u) &= \frac{f (u)}{\bar u^n}\\
F _{3n} (u) &= \frac{f (u)}{u^n}\\
\tilde{\mathcal F} _i (\alpha _i) &= \frac{\mathcal F (\alpha _i)}{y _i^n},\quad i=5,6
\end{align}
and
\begin{align}
y _5 &= \alpha _1 + \bar u \alpha _3\\
y _6 &= \alpha _1 + u \alpha _3 
\end{align}
$ u _{10} = u _{20} $ is a solution of  the equation $ m _\rho ^2 u - u q^2/\bar u = s _0 $ and $ u _{30} = u _{40} $ is a solution of the equation $ m _\rho ^2 \bar u - \bar u q^2/u = s _0 $, and finally, $ y _0 $ is a solution of the equation $ m _\rho ^2 \bar y - \bar y q ^2 / y = s _0 $. Note that in Eqs. \eqref{I1} and \eqref{I3}, the contributions of the surface terms in the leading-twist amplitudes are taken into account. 
\prg 
Defining $ Q^2 = -q^2 $, equating the corresponding coefficients of the correlation function from the QCD and hadronic parts, and solving these equations for the nine formfactors, we get
	\begin{align}
	A _0^q (Q^2) &= \frac 18 e^{\mrho^2/M^2} (\I_1[\gpera+(\gperap+4 \gperv) (-1+u),1]+\I_3[\gpera+\gperap u-4 \gperv u,1]
	\nn +4 \mrho^2 (2 \I_1[(-2 \hhgperv+\hhguc+\hhphipar) (-1+u),2]-2 \I_3[(-2 \hhgperv+\hhguc+\hhphipar) u,2]
	\nn +\I_5[(\alpha_1+\alpha_3-\alpha_3 u) (-\V+\A (-1+2 u)),2]-\I_6[(\alpha_1+\alpha_3 u) (-\V+\A (-1+2 u)),2]))\\
	A _1^q (Q^2) &= \frac{1}{Q^2} e^{\mrho^2/M^2} \mrho^2 (\I_1[\gpera+4 \hgperv-4 (\hphipar+\phipar (-1+u))+\gperap (-1+u)+4 \gperv (-1+u),1]
	\nn +\I_3[\gpera-4 \hgperv+4 \hphipar+(\gperap-4 \gperv+4 \phipar) u,1]
	\nn +\mrho^2 (-\I_1[\hApar+2 \Apar (-1+u)+4 (2 \hhgperv-\hhguc-\hhphipar+2 \hgperv (-1+u)
	\nn -2 \hphipar (-1+u)) (-1+u),2]+8 \mrho^2 (\I_1[(\hApar-2 (-2 \hhgperv+\hhguc+\hhphipar) (-1+u)) (-1+u)^2,3]
	\nn +\I_3[u^2 (-\hApar+2 (-2 \hhgperv+\hhguc+\hhphipar) u),3])
	\nn +\I_3[\hApar+2 u (\Apar+4 \hgperv u-2 (-2 \hhgperv+\hhguc+\hhphipar+2 \hphipar u)),2]
	\nn +4 \I_5[(\alpha_1+\alpha_3-\alpha_3 u) (-\V+\A (-1+2 u)),2]-4 \I_6[(\alpha_1+\alpha_3 u) (-\V+\A (-1+2 u)),2]))\\
	C _0^q (Q^2) &= \frac{1}{32\mrho^2} e^{\mrho^2/M^2} (2 (\I_1[\Apar \mrho^2-8 \gperv \mrho^2+8 \hgperv \mrho^2+4 \hhgperv \mrho^2-2 \hhguc \mrho^2-2 \hhphipar \mrho^2
	\nn -8 \hphipar \mrho^2+\gpera (2 \mrho^2-Q^2)-4 \gperv Q^2-2 \hgperv Q^2+2 \hphipar Q^2-2 \phipar (4 \mrho^2-Q^2) (-1+u)
	\nn +8 \gperv \mrho^2 u-4 \gperv Q^2 u+\gperap (2 \mrho^2 (-1+u)-Q^2 (1+u)),1]+\I_2[\gperap+2 \gperv-2 \phipar]
	\nn +\I_3[\Apar \mrho^2+\gpera (2 \mrho^2-Q^2)+\gperap (2 Q^2+2 \mrho^2 u-Q^2 u)-2 (-2 \hhgperv \mrho^2+\hhguc \mrho^2+\hhphipar \mrho^2
	\nn -4 \hphipar \mrho^2+\hgperv (4 \mrho^2-Q^2)+\hphipar Q^2-4 \mrho^2 \phipar u+\phipar Q^2 u+\gperv (-2 Q^2 (-2+u)+4 \mrho^2 u)),1]
	\nn +\I_4[-\gperap+2 \gperv-2 \phipar])+\mrho^2 (-\I_1[4 \hApar \mrho^2-\hApar Q^2-64 \hhgperv Q^2+32 \hhguc Q^2+32 \hhphipar Q^2
	\nn +8 \Apar \mrho^2 (-1+u)+32 \hhgperv \mrho^2 (-1+u)-16 \hhguc \mrho^2 (-1+u)-16 \hhphipar \mrho^2 (-1+u)
	\nn -2 \Apar Q^2 (-1+u)-24 \hhgperv Q^2 (-1+u)+12 \hhguc Q^2 (-1+u)+12 \hhphipar Q^2 (-1+u)
	\nn +32 \hgperv \mrho^2 (-1+u)^2-32 \hphipar \mrho^2 (-1+u)^2+16 \hgperv Q^2 (-1+u)^2-16 \hphipar Q^2 (-1+u)^2,2]
	\nn +16 \mrho^2 (-\I_1[(1-u) (\hApar (2 \mrho^2+Q^2) (-1+u)-2 (-2 \hhgperv+\hhguc+\hhphipar) (2 \mrho^2 (-1+u)^2
	\nn +Q^2 (-1+(-2+u) u))),3]
	\nn +\I_3[u (-\hApar (2 \mrho^2+Q^2) u+2 (-2 \hhgperv+\hhguc+\hhphipar) (-2 Q^2+(2 \mrho^2+Q^2) u^2)),3])
	\nn +\I_3[\hApar (4 \mrho^2-Q^2)+\Apar (8 \mrho^2 u-2 Q^2 u)+4 (4 \hgperv (2 \mrho^2+Q^2) u^2-4 \hphipar (2 \mrho^2+Q^2) u^2
	\nn +(-2 \hhgperv+\hhguc+\hhphipar) (-4 \mrho^2 u+Q^2 (-8+3 u))),2]
	\nn +4 \I_5[-\V (4 Q^2+(\alpha_1+\alpha_3-\alpha_3 u) (4 \mrho^2+Q^2-6 Q^2 u))+\A (-4 Q^2+8 Q^2 u-\alpha_1 (Q^2+\mrho^2 (4-8 u)
	\nn +4 Q^2 u)+\alpha_3 (-1+u) (Q^2+\mrho^2 (4-8 u)+4 Q^2 u)),2]-4 \I_6[\A (-4 Q^2+\alpha_1 Q^2 (5-4 u)+8 Q^2 u
	\nn +\alpha_3 Q^2 (5-4 u) u+\alpha_1 \mrho^2 (-4+8 u)+\alpha_3 \mrho^2 u (-4+8 u))-\V (4 Q^2
	\nn +(\alpha_1+\alpha_3 u) (4 \mrho^2+Q^2 (-5+6 u))),2]))\\
	C _1^q (Q^2) &= -\frac{1}{2Q^2} e^{\mrho^2/M^2} (\I_1[\gpera (2 \mrho^2-Q^2)+\gperap (4 \mrho^2 (-1+u)-Q^2 (1+u))+4 (\mrho^2 (4 \hgperv+2 \hhgperv
	\nn -\hhguc-\hhphipar-4 \hphipar-4 \phipar (-1+u))+\gperv (4 \mrho^2 (-1+u)-Q^2 (1+u))),1]+\I_2[\gperap]
	\nn +\I_3[\gpera (2 \mrho^2-Q^2)+4 \gperv (Q^2 (-2+u)-4 \mrho^2 u)-4 \mrho^2 (4 \hgperv-2 \hhgperv+\hhguc+\hhphipar-4 \hphipar
	\nn -4 \phipar u)+\gperap (2 Q^2+4 \mrho^2 u-Q^2 u),1]-\I_4[\gperap]+2 \mrho^2 (\I_1[-\hApar \mrho^2-2 \gpera Q^2
	\nn -4 \Apar \mrho^2 (-1+u)+4 (-2 \hhgperv+\hhguc+\hhphipar) (\mrho^2 (-1+u)-Q^2 (1+u))-4 \hgperv (4 \mrho^2 (-1+u)^2
	\nn +Q^2 (3+(-2+u) u))+4 \hphipar (4 \mrho^2 (-1+u)^2+Q^2 (3+(-2+u) u)),2]+4 \mrho^2 (\I_1[-2 (-2 \hhgperv+\hhguc
	\nn +\hhphipar) (4 \mrho^2+Q^2) (-1+u)^3+\hApar (4 \mrho^2 (-1+u)^2+Q^2 (3+(-2+u) u)),3]-\I_3[-2 (-2 \hhgperv\nn +\hhguc+\hhphipar) (4 \mrho^2+Q^2) u^3+\hApar (2 Q^2+(4 \mrho^2+Q^2) u^2),3])+\I_3[\hApar \mrho^2-2 (\gpera-4 \hgperv) Q^2
	\nn +4 \Apar \mrho^2 u+4 \hgperv (4 \mrho^2+Q^2) u^2+4 (-2 \hhgperv+\hhguc+\hhphipar) (Q^2 (-2+u)-\mrho^2 u)-4 \hphipar (2 Q^2
	\nn +(4 \mrho^2+Q^2) u^2),2]+4 \I_5[-\V (Q^2+(\alpha_1+\alpha_3-\alpha_3 u) (2 \mrho^2-Q^2 u))-\A (Q^2-2 Q^2 u
	\nn +\alpha_1 (\mrho^2 (2-4 u)+Q^2 u)+\alpha_3 (-1+u) (-Q^2 u+\mrho^2 (-2+4 u))),2]-4 \I_6[-\V (Q^2+(2 \mrho^2
	\nn +Q^2 (-1+u)) (\alpha_1+\alpha_3 u))+\A (-Q^2+2 Q^2 u+\alpha_1 (Q^2-Q^2 u+\mrho^2 (-2+4 u))+\alpha_3 u (Q^2-Q^2 u
	\nn +\mrho^2 (-2+4 u))),2]))\\
	D _0^q (Q^2) &= \frac 18 e^{\mrho^2/M^2} (\I_1[\gpera+(\gperap+4 \gperv) (1+u),1]+\I_3[\gpera-(-\gperap+4 \gperv) (-2+u),1]
	\nn +4 \mrho^2 (2 \I_1[(-2 \hhgperv+\hhguc+\hhphipar) (1+u),2]-2 \I_3[(-2 \hhgperv+\hhguc+\hhphipar) (-2+u),2]
	\nn +\I_5[(-2+\alpha_1+\alpha_3-\alpha_3 u) (-\V+\A (-1+2 u)),2]
	\nn -\I_6[(-2+\alpha_1+\alpha_3 u) (-\V+\A (-1+2 u)),2]))\\
	D _1^q (Q^2) &= -\frac{1}{Q^4} e^{\mrho^2/M^2} \mrho^2 (-\I_1[\gpera (8 \mrho^2-Q^2)+\gperap (8 \mrho^2 (-1+u)-Q^2 (1+u))+4 ((8 \mrho^2-Q^2) (\hgperv
	\nn -\hphipar+\phipar-\phipar u)+\gperv (8 \mrho^2 (-1+u)-Q^2 (1+u))),1]-\I_3[\gpera (8 \mrho^2-Q^2)
	\nn +4 \gperv (Q^2 (-2+u)-8 \mrho^2 u)-4 (8 \mrho^2-Q^2) (\hgperv-\hphipar-\phipar u)+\gperap (2 Q^2+8 \mrho^2 u-Q^2 u),1]
	\nn +\mrho^2 (\I_1[8 \hApar \mrho^2+4 \gpera Q^2-\hApar Q^2+16 \hgperv Q^2-32 \hhgperv Q^2+16 \hhguc Q^2+16 \hhphipar Q^2
	\nn -16 \hphipar Q^2+16 \Apar \mrho^2 (-1+u)+64 \hhgperv \mrho^2 (-1+u)-32 \hhguc \mrho^2 (-1+u)-32 \hhphipar \mrho^2 (-1+u)
	\nn -2 \Apar Q^2 (-1+u)-8 \hhgperv Q^2 (-1+u)+4 \hhguc Q^2 (-1+u)+4 \hhphipar Q^2 (-1+u)
	\nn +64 \hgperv \mrho^2 (-1+u)^2-64 \hphipar \mrho^2 (-1+u)^2+24 \hgperv Q^2 (-1+u)^2-24 \hphipar Q^2 (-1+u)^2,2]
	\nn +\I_3[-8 \hApar \mrho^2+4 \gpera Q^2+\hApar Q^2-16 \hgperv Q^2-32 \hhgperv Q^2+16 \hhguc Q^2+16 \hphipar Q^2
	\nn -64 \hhgperv \mrho^2 u+32 \hhguc \mrho^2 u+8 \hhgperv Q^2 u-4 \hhguc Q^2 u+2 \Apar (-8 \mrho^2+Q^2) u-64 \hgperv \mrho^2 u^2
	\nn +64 \hphipar \mrho^2 u^2-24 \hgperv Q^2 u^2+24 \hphipar Q^2 u^2+4 \hhphipar (4 Q^2+8 \mrho^2 u-Q^2 u),2]+8 \mrho^2 (\I_1[2 (-2 \hhgperv
	\nn +\hhguc+\hhphipar) (-1+u) (8 \mrho^2 (-1+u)^2+Q^2 (-1+3 (-2+u) u))-\hApar (8 \mrho^2 (-1+u)^2
	\nn +Q^2 (5+3 (-2+u) u)),3]+\I_3[-2 (-2 \hhgperv+\hhguc+\hhphipar) u (-4 Q^2+(8 \mrho^2+3 Q^2) u^2)+\hApar (2 Q^2
	\nn +(8 \mrho^2+3 Q^2) u^2),3])-4 \I_5[\A (-2 Q^2-(\alpha_1+\alpha_3) (8 \mrho^2+3 Q^2)+8 (2 \alpha_1+3 \alpha_3) \mrho^2 u
	\nn +(4-2 \alpha_1+\alpha_3) Q^2 u+2 \alpha_3 (-8 \mrho^2+Q^2) u^2)-\V (2 Q^2+(8 \mrho^2+Q^2 (3-8 u)) (\alpha_1+\alpha_3-\alpha_3 u)),2]
	\nn +4 \I_6[\A (\alpha_1 Q^2 (5-2 u)+8 \alpha_1 \mrho^2 (-1+2 u)+8 \alpha_3 \mrho^2 u (-1+2 u)+Q^2 (-2+(4+\alpha_3 (5-2 u)) u))
	\nn -\V (2 Q^2+(\alpha_1+\alpha_3 u) (8 \mrho^2+Q^2 (-5+8 u))),2]))\\
	E ^q (Q^2)   &= \frac{1}{8Q^2} e^{\mrho^2/M^2} (4 (\I_1[\gpera \mrho^2-4 \hphipar \mrho^2+\hgperv (4 \mrho^2-Q^2)+\hphipar Q^2-\gperap \mrho^2 (1-u)
	\nn -4 \gperv \mrho^2 (1-u)+4 \mrho^2 \phipar (1-u)-\phipar Q^2 (1-u),1]+\I_3[\gpera \mrho^2+\gperap \mrho^2 u-4 \gperv \mrho^2 u
	\nn -(4 \mrho^2-Q^2) (\hgperv-\hphipar-\phipar u),1])+\mrho^2 (-\I_1[\hApar (4 \mrho^2-Q^2)+2 (1-u) (\Apar (-4 \mrho^2+Q^2)
	\nn +2 (-2 \hhgperv+\hhguc+\hhphipar) (4 \mrho^2+Q^2)+8 (\hgperv-\hphipar) (2 \mrho^2+Q^2) (1-u)),2]
	\nn +\I_3[\hApar (4 \mrho^2-Q^2)+\Apar (8 \mrho^2 u-2 Q^2 u)+4 u (-(-2 \hhgperv+\hhguc+\hhphipar) (4 \mrho^2+Q^2)
	\nn +4 \hgperv (2 \mrho^2+Q^2) u-4 \hphipar (2 \mrho^2+Q^2) u),2]+16 \mrho^2 (\I_1[(1-u) (-4 (-2 \hhgperv+\hhguc+\hhphipar) Q^2
	\nn +\hApar (2 \mrho^2+Q^2) (1-u)+2 (-2 \hhgperv+\hhguc+\hhphipar) (2 \mrho^2+Q^2) (1-u)^2),3]
	\nn +\I_3[u (-\hApar (2 \mrho^2+Q^2) u+2 (-2 \hhgperv+\hhguc+\hhphipar) (-2 Q^2+(2 \mrho^2+Q^2) u^2)),3])
	\nn -4 (\I_5[(\alpha_1+\alpha_3 (1-u)) (\V (4 \mrho^2 u-3 Q^2 u+4 \mrho^2 (1-u)+3 Q^2 (1-u))+\A (-4 \mrho^2 u+3 Q^2 u
	\nn +4 \mrho^2 (1-u)+3 Q^2 (1-u))),2]+\I_6[(\alpha_1+\alpha_3 u) (-\V (4 \mrho^2 u+3 Q^2 u+4 \mrho^2 (1-u)-3 Q^2 (1-u))
	\nn +\A (4 \mrho^2 u+3 Q^2 u-4 \mrho^2 (1-u)+3 Q^2 (1-u))),2])))\\
	F ^q (Q^2)   &= \frac{1}{16\mrho^2} e^{\mrho^2/M^2} (2 (\I_1[\Apar \mrho^2+2 (\gpera \mrho^2-2 \hhgperv \mrho^2+\hhguc \mrho^2+\hhphipar \mrho^2-4 \hphipar \mrho^2
	\nn +\hgperv (4 \mrho^2-Q^2)+\hphipar Q^2-\gperap \mrho^2 (1-u)-4 \gperv \mrho^2 (1-u)+4 \mrho^2 \phipar (1-u)
	\nn -\phipar Q^2 (1-u)),1]+2 \I_2[\gperv-\phipar]+\I_3[\Apar \mrho^2+2 (\gpera \mrho^2-4 \hgperv \mrho^2-2 \hhgperv \mrho^2
	\nn +\hhguc \mrho^2+\hhphipar \mrho^2+4 \hphipar \mrho^2+\hgperv Q^2-\hphipar Q^2+\gperap \mrho^2 u-4 \gperv \mrho^2 u+4 \mrho^2 \phipar u
	\nn -\phipar Q^2 u),1]+2 \I_4[\gperv-\phipar])+\mrho^2 (-\I_1[\hApar (4 \mrho^2-Q^2)+2 (1-u) (\Apar (-4 \mrho^2+Q^2)
	\nn +2 (-2 \hhgperv+\hhguc+\hhphipar) (4 \mrho^2+Q^2)+8 (\hgperv-\hphipar) (2 \mrho^2+Q^2) (1-u)),2]
	\nn +\I_3[\hApar (4 \mrho^2-Q^2)+\Apar (8 \mrho^2 u-2 Q^2 u)+4 u (-(-2 \hhgperv+\hhguc+\hhphipar) (4 \mrho^2+Q^2)
	\nn +4 \hgperv (2 \mrho^2+Q^2) u-4 \hphipar (2 \mrho^2+Q^2) u),2]+16 \mrho^2 (\I_1[(1-u) (-4 (-2 \hhgperv+\hhguc+\hhphipar) Q^2
	\nn +\hApar (2 \mrho^2+Q^2) (1-u)+2 (-2 \hhgperv+\hhguc+\hhphipar) (2 \mrho^2+Q^2) (1-u)^2),3]
	\nn +\I_3[u (-\hApar (2 \mrho^2+Q^2) u+2 (-2 \hhgperv+\hhguc+\hhphipar) (-2 Q^2+(2 \mrho^2+Q^2) u^2)),3])
	\nn -4 (\I_5[(\alpha_1+\alpha_3 (1-u)) (\V (4 \mrho^2 u-3 Q^2 u+4 \mrho^2 (1-u)+3 Q^2 (1-u))+\A (-4 \mrho^2 u+3 Q^2 u
	\nn +4 \mrho^2 (1-u)+3 Q^2 (1-u))),2]+\I_6[(\alpha_1+\alpha_3 u) (-\V (4 \mrho^2 u+3 Q^2 u+4 \mrho^2 (1-u)-3 Q^2 (1-u))
	\nn +\A (4 \mrho^2 u+3 Q^2 u-4 \mrho^2 (1-u)+3 Q^2 (1-u))),2])))\\
	J ^q (Q^2)   &= \frac 18 e^{\mrho^2/M^2} (4 (\I_1[-\hgperv+\hphipar+\phipar (-1+u),1]-\I_3[-\hgperv+\hphipar+\phipar u,1])+\mrho^2 (\I_1[\hApar
	\nn +2 \Apar (-1+u)+4 (-2 \hhgperv+\hhguc+\hhphipar) (-1+u),2]-\I_3[\hApar+2 \Apar u+4 (-2 \hhgperv+\hhguc
	\nn +\hhphipar) u,2]-4 (\I_5[(\alpha_1+\alpha_3-\alpha_3 u) (\A+\V-2 \V u),2]+\I_6[(\alpha_1+\alpha_3 u) (\A+\V
	\nn -2 \V u),2])))
	\end{align}

\section{Numerical analysis}
\begin{figure}
	\centering 
	\includegraphics[width=.5\linewidth]{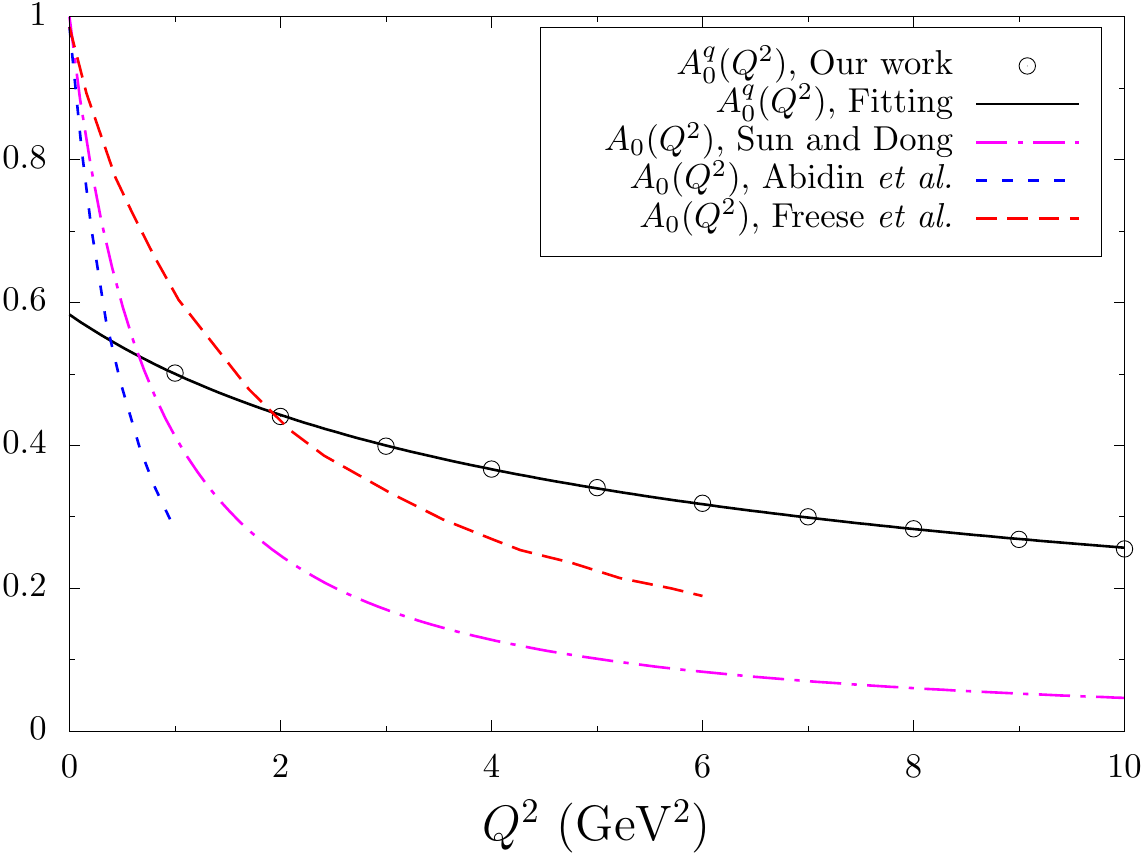}\caption{The GFFs of the $ \rho $ meson: $ A _0^q (Q^2) $. The results of Sun and Dong \cite{ref:11} in the light-cone constituent quark model, Abidin \textit{et al.} \cite{ref:19} from the AdS/QCD approach, and Freese \textit{et al.} \cite{ref:21} in the NJL model are also shown.}\label{fig:1}
\end{figure}
\begin{figure}
	\centering 
	\includegraphics[width=.5\linewidth]{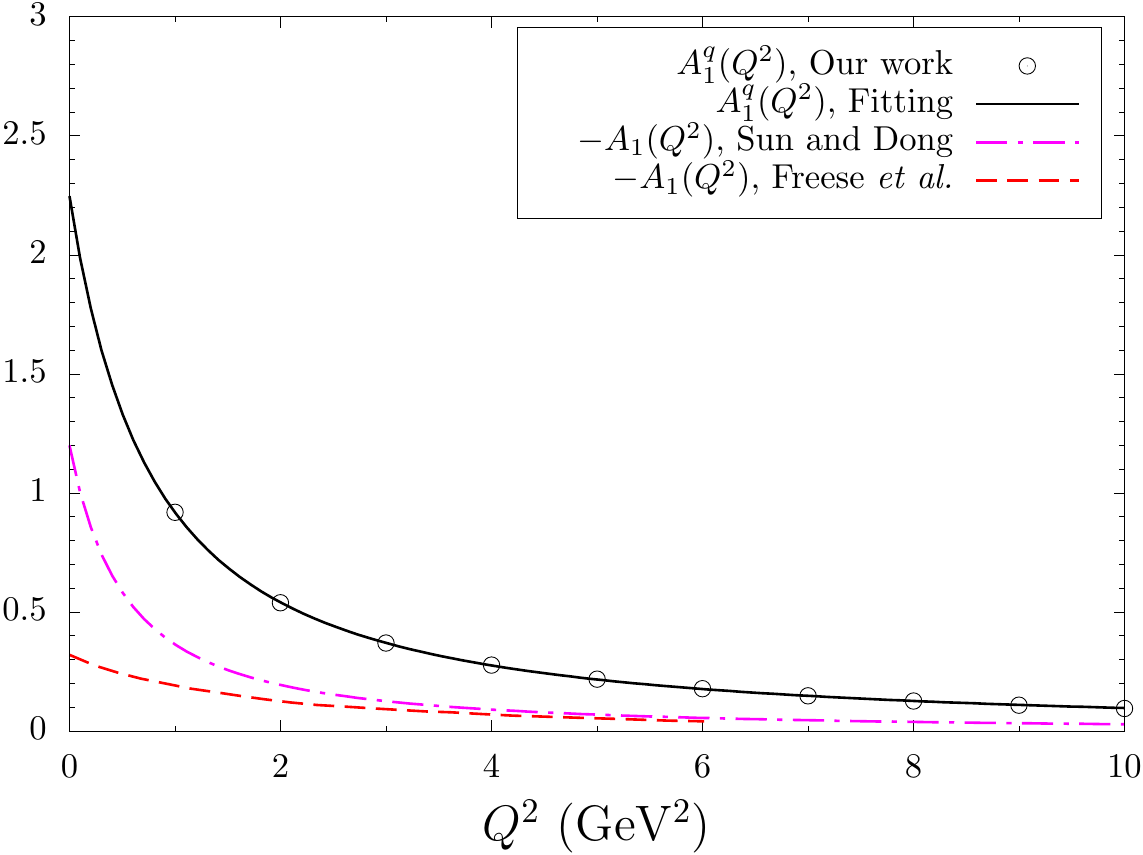}\caption{The GFFs of the $ \rho $ meson: $ A _1^q (Q^2) $. The results of Sun and Dong \cite{ref:11} in the light-cone constituent quark model and Freese \textit{et al.} \cite{ref:21} in the NJL model are also shown.}\label{fig:2}
\end{figure}
\begin{figure}
	\centering 
	\includegraphics[width=.5\linewidth]{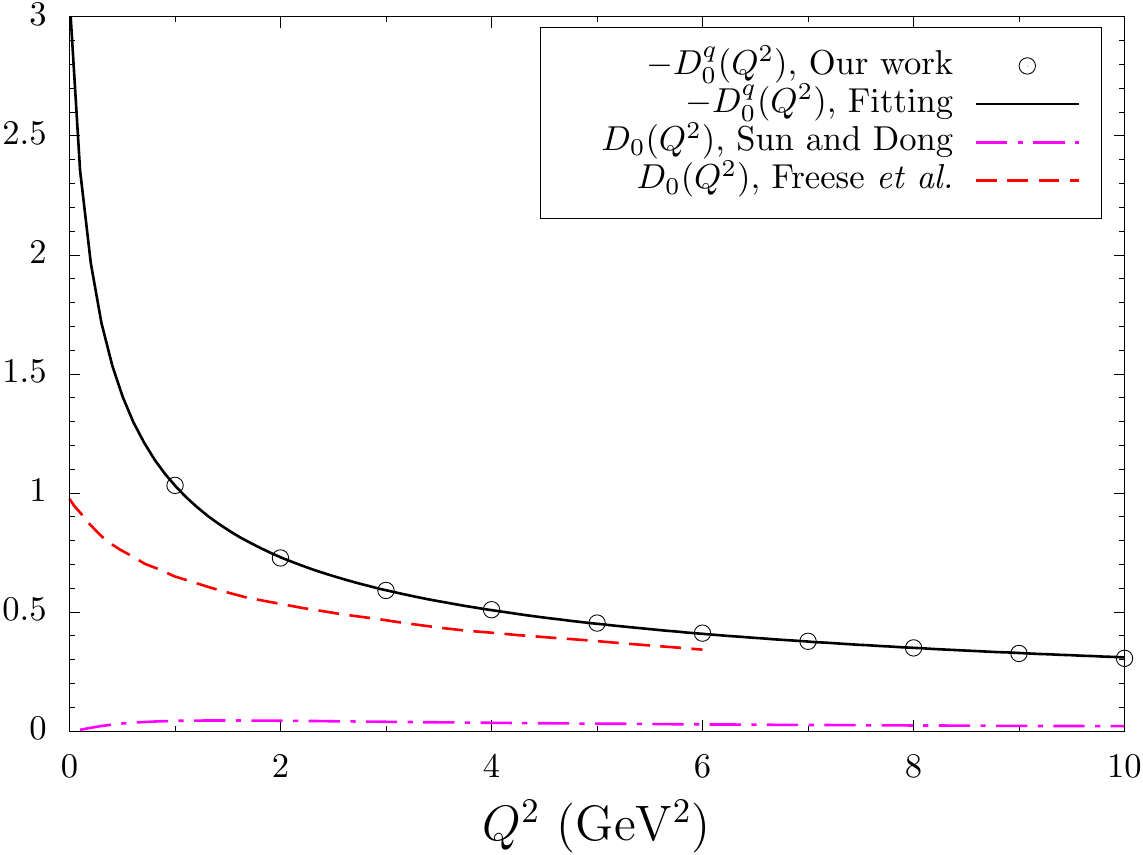}\caption{The GFFs of the $ \rho $ meson: $ D _0^q (Q^2) $. The results of Sun and Dong \cite{ref:11} in the light-cone constituent quark model and Freese \textit{et al.} \cite{ref:21} in the NJL model are also shown.}\label{fig:3}
\end{figure}
\begin{figure}
	\centering 
	\includegraphics[width=.5\linewidth]{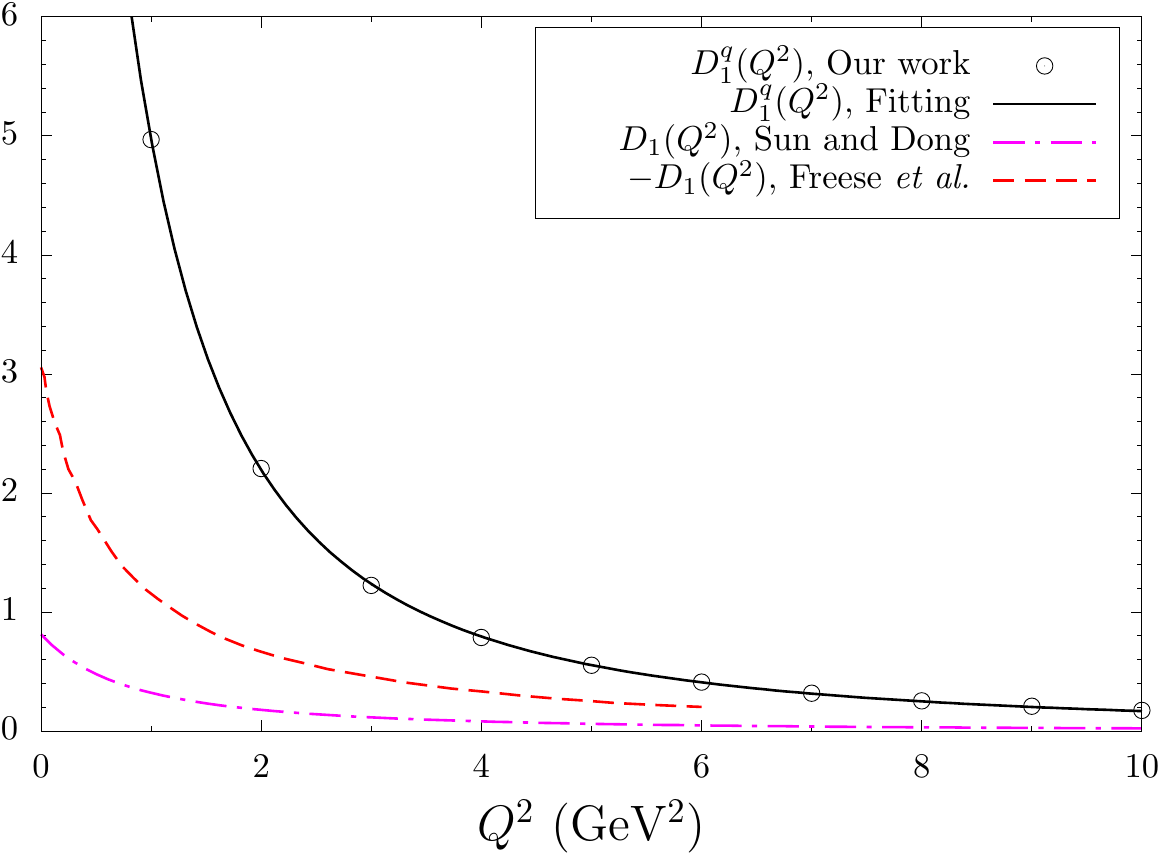}\caption{The GFFs of the $ \rho $ meson: $ D _1^q (Q^2) $. The results of Sun and Dong \cite{ref:11} in the light-cone constituent quark model and Freese \textit{et al.} \cite{ref:21} in the NJL model are also shown.}\label{fig:4}
\end{figure}
\begin{figure}
	\centering 
	\includegraphics[width=.5\linewidth]{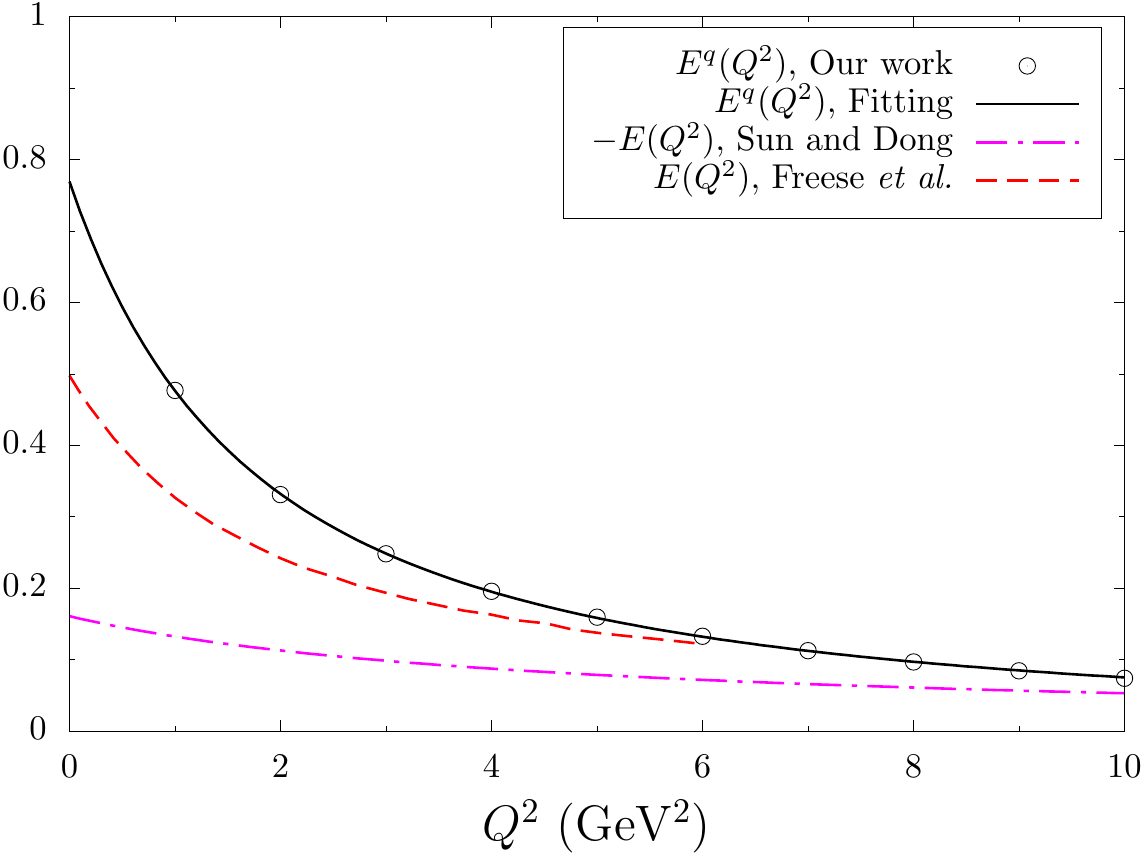}\caption{The GFFs of the $ \rho $ meson: $ E^q (Q^2) $. The results of Sun and Dong \cite{ref:11} in the light-cone constituent quark model and Freese \textit{et al.} \cite{ref:21} in the NJL model are also shown.}\label{fig:5}
\end{figure}
\begin{figure}
	\centering 
	\includegraphics[width=.5\linewidth]{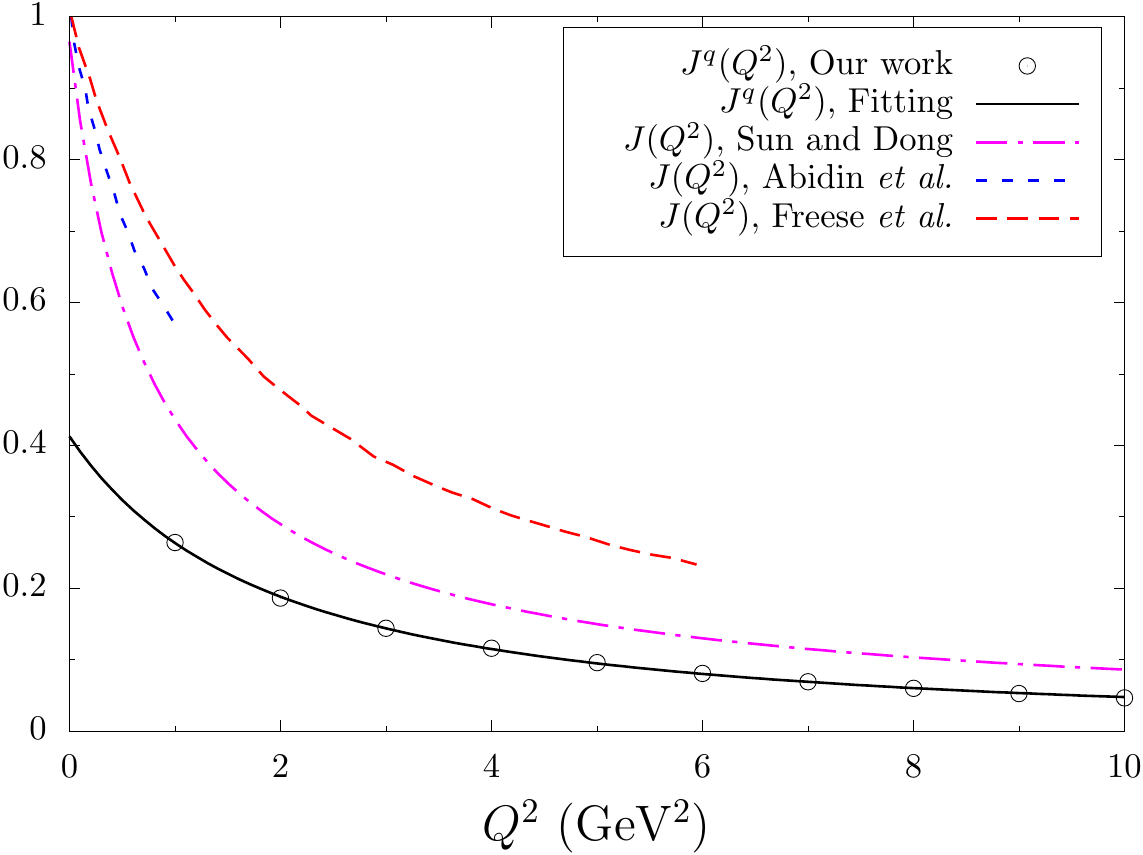}\caption{The GFFs of the $ \rho $ meson: $ J^q (Q^2) $. The results of Sun and Dong \cite{ref:11} in the light-cone constituent quark model, Abidin \textit{et al.} \cite{ref:19} from the AdS/QCD approach, and Freese \textit{et al.} \cite{ref:21} in the NJL model are also shown.}\label{fig:6}
\end{figure}
In this section, we numerically analyze the light-cone sum rules for the GFFs of the $ \rho $ meson by using Package X \cite{ref:X}. In the sum rules, we took the mass and decay constant of the $ \rho $ meson to be $ m _\rho = 0.77 {\rm\ GeV} $ and $ f _\rho = 0.20 {\rm\ GeV}$, respectively. Another set of essential input parameters are the $ \rho $ meson DAs of different twists. The relevant DAs are given as follows \cite{ref:14, ref:15, ref:15.1, ref:16}:
\begin{align}
\phipar &= 6u \bar u \Big[
1 + 3 a _1 ^\parallel \xi + a _2^\parallel \frac 32 (5\xi^2-1)
\Big]\\
\guc &= 1 + \Big(
-1 - \frac 27 a _2^\parallel + \frac{40}{3} \zeta _3^A - \frac{20}{3} \zeta _4
\Big) C _2^{1/2} (\xi) 
 + \Big[
-\frac{27}{28} a _2^\parallel + \frac{5}{4} \zeta _3^A - \frac{15}{16} \zeta _3^A (\omega _3 ^A + 3 \omega _3 ^V) 
\Big] C _4^{1/2} (\xi) \\
\gpera &= 6u\bar u \Big\{
1 + a _1^\parallel \xi + \Big[
\frac 14 a _2 ^\parallel + \frac 53 \zeta _3^A \Big(
1- \frac{3}{16} \omega _3 ^A
\Big) 
 + \frac{35}{4} \zeta _3^V
\Big] (5\xi^2 -1)
\}\\
\gperv &= \frac 34 (1+\xi^2) + a _1^\parallel \frac 32 \xi ^3 + \Big(
\frac 37 a _2^\parallel + 5 \zeta _3 ^A
\Big) (3\xi^2 - 1) 
 + \Big(
\frac{9}{112} a _2 ^\parallel + \frac{105}{16} \zeta _3 ^V - \frac{15}{64} \zeta _3 ^A \omega _3 ^A
\Big)   (3-30\xi^2+35\xi^4) \\
\Apar &= 24 u^2 \bar u ^2\\
\V &= 5040 (\alpha _1 - \alpha _2) \alpha _1 \alpha _2 \alpha _3^2\\
\A &= 360 \alpha _1 \alpha _2 \alpha _3 ^2 \Big[
1 + \omega _3^A \frac 12 (7\alpha _3 - 3)
\Big]
\end{align}
The $ C _n^k(x) $ are the Gegenbauer polynomials, $ \bar u = 1-u $, and $ \xi = 2u-1 $. The values of the parameters inside the DAs at the renormalization scale of $ \mu = 1 {\rm\ GeV} $ are $ a _1^\parallel = 0 $, $ a _2 ^\parallel = 0.18 $, $ \zeta _3 ^A = 0.032 $, $ \zeta _4 = 0.15 $, $ \omega _3^A = -2.1 $, $ \omega _3 ^V = 3.8 $, and $ \zeta _3^V = 0.013 $.
\prg
From the sum rules for the GFFs, we see that besides the input parameters, they contain two auxiliary parameters, namely the Borel mass parameter, $ M^2 $, and the continuum threshold, $ s _0 $. Apparently, the measurable GFFs should be independent of them. We find the working region of $ M^2 $ to be 
\begin{align}
1.0 {\rm\ GeV} < M^2 < 2.0 {\rm\ GeV}
\end{align}
and the continuum threshold to be 
\begin{align}
s _0 = 1.4 {\rm\ GeV^2}
\end{align}
We present our results for the six GFFs that lead to the conservation of the EMT in Figs. \ref{fig:1}--\ref{fig:6}. The GFFs $ A _0 (q^2) $ and $ J(q^2) $ are related to the mass and charge conservation, hence are subject to the constraint at zero-momentum transfer $ A _0(0) = 1 $ and $ J(0) = 1 $ \cite{ref:17, ref:18, ref:19, ref:20}. It is crucial to note that in this work, we took into account only the contribution from the quarks in the EMT. 
\prg 
As we noted that, the correlation function from QCD side can be calculated at sufficiently large negative values of $ Q^2 $. The formfactors can be reliably determined at $ Q^2 \geq 1 {\rm\ GeV^2} $ domain. The LCSR method is not applicable for smaller values of $ Q^2 $: $ Q^2 <1 {\rm\ GeV^2} $. In order to extend the results for the formfactors to $ Q^2 = 0 $ point, we look for a parametrization of them in such a way that at large $ Q^2 $ domain, the parametrization coincides with the sum rules predictions. Our numerical analysis shows that the best parametrization for the formfactors is as follows:
\begin{align}
A _0^q (Q^2) &= 0.583 \Big(1 + \frac{Q^2}{3.451}
\Big) ^ {-0.603}\\
A _1^q(Q^2) &= 2.247 \Big(1 + \frac{Q^2}{1.042}\Big) ^{-1.329}\\
D _0^q(Q^2) &= -3.086 \Big(1 + \frac{Q^2}{0.160}\Big)^{-0.554}\\
D _1^q(Q^2) &= 21.04 \Big(1+\frac{Q^2}{0.904}\Big)^{-1.936}\\
E^q(Q^2) &= 0.769 \Big(1+\frac{Q^2}{2.603}\Big)^{-1.473}\\
J^q(Q^2) &= 0.413 \Big(1+\frac{Q^2}{2.541}\Big)^{-1.351}
\end{align}
From Figs. \ref{fig:1}--\ref{fig:6}, we deduce the following results: The contributions from the quark sector of the EMT to the GFF for $ A _0^q $ seem to make up nearly 60\% of the total contributions. For the GFF $ J $, 40\% of the total contribution consists of the quark part. The results for both formfactors agree with the expectation that the quark sector in the EMT will add up to almost half of the total contributions to the GFFs. For the rest of the GFFs displayed in Figs. \ref{fig:1}--\ref{fig:6}, we see that the quark contributions seem to be larger than the total obtained in the literature, which would indicate that the gluon contributions should have largely negative contributions.
\prg 
At the end of this section, we present our predictions for the average mass radius, $ \langle r _{\rm mass}^2 \rangle $, and the quadrupole moment of the $ \rho $ meson. The average mass radius $ \langle r^2 \rangle _{\rm mass} $ is obtained in the light-cone frame in Ref. \cite{ref:21} as 
\begin{align}
\langle r^2 \rangle _{\rm mass} &= 4 \frac{dA _0(q^2)}{dq^2} \Big | _{q^2 = 0}
 + \frac{1}{3m^2} [2A _0(0) + A _1(0) - 2 J(0) +2E(0)]
\end{align}
In our work, we obtain
\begin{align}
\sqrt{ \langle r^2 \rangle _{\rm mass} } = 0.32 {\rm\ fm}
\end{align}
while it was found to be $ 0.41 {\rm\ fm} $ in \cite{ref:11}. Finally, the gravitational quadrupole moment is given in terms of the GFFs as [18]
\begin{align}
\mathcal Q _{\rm mass} &= - \frac{1}{m} \Big[
- A _0 (0) + \frac 12 A _1 (0) + 2 J (0) - E (0)
\Big]
\end{align}
In our work, we obtain
\begin{align}
\mathcal Q _{\rm mass} = -0.0512 \ \mrho \cdot {\rm fm^2}
\end{align}
while it was obtained to be $ -0.0322 \ \mrho \cdot {\rm fm^2} $ in \cite{ref:11}. 
\section{Conclusion}
In this work, we studied the GFFs of the $ \rho $ meson within the light-cone QCD sum rules approach by taking into account only the quark part in the EMT. The GFFs $ A _0 $ and $ J $ are related to the mass and charge conservation and they are subject to the constraint $ A _0(0) = 1 $ and $ J (0) = 1 $ at zero transfer momentum. We have shown that the quark contributions make up nearly 60\% and 40\% of the aforementioned GFFs, respectively. The mass radius was obtained to be $ 0.32 {\rm\ fm}$ and the gravitational quadrupole moment of the $ \rho $ meson was found to be $ -0.0512 \ \mrho \cdot {\rm fm^2} $. Finally, we compared our results to the ones in the literature, and considering the fact that the gluon part has been neglected in the EMT, the results do not differ significantly. 
\section*{Acknowledgments}
One of the authors (T. Barakat) extends his appreciation to the Deanship Scientific Research at King Saud University for funding his work through research program RG-1440-090.
\bibliographystyle{ieeetr}
\bibliography{submit_to_arxiv}
\end{document}